\documentclass[aps,prd,groupedaddress,showpacs]{revtex4}
\usepackage{graphicx,epsf,amssymb,amsmath,latexsym}

\newcommand{\be}{\begin{eqnarray}}
\newcommand{\ee}{\end{eqnarray}}

\def\ben{\begin{equation}}
\def\een{\end{equation}}
\def\bena{\begin{eqnarray}}
\def\eena{\end{eqnarray}}

\begin{document}

\title{The Cosmological Slingshot Scenario in details}

\author{Cristiano Germani}
\email{germani@sissa.it}
\affiliation{SISSA and INFN, via Beirut 4, 34014 Trieste, Italy}
\author{Nicol\'as E. Grandi}
\email{grandi@fisica.unlp.edu.ar}
\affiliation{IFLP-CONICET, cc67, CP1900 La Plata, Argentina}
\author{Alex Kehagias}
\affiliation{Physics Division, National Technical University of Athens,
\\ 15780 Zografou Campus,  Athens, Greece}
\email{kehagias@central.ntua.gr}

\begin{abstract}
We generalize  the Cosmological Slingshot Scenario for a Slingshot
brane moving in a Klebanov-Strassler throat.
We show that the horizon and isotropy problems of standard cosmology are avoided,
while the flatness problem  is acceptably alleviated. Regarding the
primordial perturbations, we identify their vacuum state and
elucidate the evolution from the quantum to the classical regimes.
Also, we calculate their exact power spectrum showing its
compatibility with current data. We discuss the bouncing
solution from a four dimensional point of view.
In this framework the radial and angular motion of the Slingshot brane are described by two scalar fields.
We show that the bouncing solution for the scale factor in String frame is mapped into a monotonically
increasing (in conformal time) solution
in the Einstein frame. We finally discuss about the regularity of the geometry in Einstein frame.
.
\end{abstract}

\pacs{}

\vskip1.5cm

{\hfill {SISSA 32/2007/A}}

\maketitle

\section{Introduction}

The Cosmological Slingshot Scenario \cite{sling}, shortly the
``Slingshot", is a proposal for the cosmic early-time evolution in
the String Theory context. According to that, our Universe is a
$D3$-brane moving in a String Theory background of the form
${\cal M}^4\times K^6$. ${\cal M}^4$ is a ``warped" Minkowskian
space-time and $K^6$ is a compact Calabi-Yau (CY) space. The
latter includes a ``throat" sourced by a stack of a large number
($N$) of other $D3$ branes. The Slingshot is characterized by a
non-trivial orbital motion of the Universe in the compact space
around the stack of $D3$-branes. If back-reactions due to the
brane motion can be neglected (probe brane approximation), a brane
observer measures a $3+1$ dimensional induced metric in terms of
the brane embedding. This metric defines a cosmological brane
evolution commonly called Mirage Cosmology \cite{AK}.

The early-time evolution ({\it i.e.} well before nucleosynthesis)
corresponds to the motion of the $D3$-brane deep into the throat (Slingshot era) moving towards
the hat of the compact space. The
late-time cosmology starts when the $D3$-brane reaches the hat of the CY. There,
the probe brane approximation breaks down and local gravity
{\it \`a la} Randall-Sundrum \cite{rs} dominates the cosmological evolution.
Since $N$ is taken to be large, close to the
stack, the probe brane approximation of \cite{AK} can be used. Under this approximation, the
Slingshot brane observer experiences, at early-times, a non-singular bouncing
cosmology
in the String frame. In Einstein frame on the other hand, the induced cosmology is monotonically increasing (in
conformal time). Moreover, assuming
that in the past infinity the brane starts in the hat of the Calabi-Yau without modifying its regular structure,
no past singularities can ever be present in the both frames, as we shall argue in the paper. This scenario
is then a realization of the emergent Universe idea \cite{EM}, although in a technically very different way.
Additionally, as we shall show later on, the brane induced cosmology naturally avoids the
Standard Cosmology problems ({\it i.e.} horizon, isotropy and flatness).

This model can be included in a more general class of bouncing/cyclic models that have successfully tried to address some of the Standard
Cosmology problems, such that ekpyrotic models \cite{Khoury}, phantom based cyclic models \cite{Brown} and emergent cyclic models \cite{Biswas}.
For a review on bouncing cosmologies see \cite{santiago}.

The plan of the paper is as follows. In section \ref{sKS} we show how the standard cosmological problems are solved
when the slingshot brane
is moving in a Klebanov-Strassler throat. In section \ref{sPert} we calculate the primordial perturbation spectrum and the
corresponding spectral index. In section \ref{sEff} we write the effective 4D theory, which
 reproduces in the string frame the mirage solution, we discuss the Slingshot in the Einstein frame and we finally conclude in  section \ref{sConcl}.

\section{The Cosmological Slingshot Scenario in a Klebanov-Strassler throat}
\label{sKS}
To make the discussion concrete, we shall consider a probe $D3$-brane moving in the Klebanov-Strassler
(KS) throat \cite{KS} of a Calabi-Yau (CY)
compact manifold. The KS geometry is obtained by putting together a
stack of $N$ $D3$-branes and $M$ fractional branes at the apex of a
conifold \cite{C1,C2,C3,C4}. Then, the conical singularity is
deformed by blowing up a $3$-sphere at the tip \cite{CO}. We
restrict the probe brane motion to a region far from the tip of the
KS geometry. In this case, the KS geometry can be well approximated
by the Klebanov-Tseytlin (KT) metric \cite{KT}
\bena
ds^2=h^{-\frac{1}{2}}ds^2_{\mbox{\tiny Mink}} +
h^{\frac{1}{2}}\left(dr^2+r^2 ds_{\mathbb{T}^{1,1}}^2\right)\ ,
\label{metrik}
\eena
where $ds_{\mathbb{T}^{1,1}}^2$ is the metric of a
$\mathbb{T}^{1,1}$ manifold. The warp factor in the metric (\ref{metrik}) is
\be
h=\frac{L^4\ln(r/r_s)}{r^4}\, , ~~~L^2=9/(2\sqrt2)
l_s^2 M g_s\, ,
\ee
where $r_s$ is proportional to  the radius of the blown up sphere,   $l_s$ is the string length and $g_s$  is the
string coupling. The supergravity approximation is
valid as long as the curvature radius of the solution is large
compared to the string length $l_s$. String perturbation theory
on the other hand requires $g_s\ll 1$.

In what follows, we will assume that moduli are stabilized by some mechanism that does  not
affect the motion of our probe brane. Since we are assuming that back-reaction effects
are negligible, we can expand any potential arising from moduli stabilization in powers of the
probe brane tension in units of local curvatures. In this way, the general results of \cite{Baumann:2006th}
for stabilization mechanisms that include Euclidean $D3$-branes wrapping four cycles in the
CY, or non-perturbative effects on $D7$-branes, imply no modification for our probe brane
potential to first order.

The dynamics of a probe brane is governed by the Dirac-Born-Infeld
action with a Wess-Zumino term (that takes into account the coupling
to the bulk Ramond-Ramond five-form with the brane charge):
\ben S_{DBI}+S_{WZ}= -T_3\int \sqrt{-g_i}\,d^4\xi - T_3\int C_{(4)}
\, . \label{actionn}
\een
In the KS background $C_{0123}=1-1/h$. We assume that
all other fields on the brane are switched off and matter is
created later.
The sign of the Wess-Zumino term has been chosen to
represent a $D3$-brane in the mostly plus convention for the
metric, and $T_3=1/{(2\pi)^3g_sl_s^4}$ is the tension of the
probe.
The probe brane is extended
parallel to the stack of $D3$s, so that it looks like a point particle
moving in transverse space (for inhomogeneous embedding see
\cite{cristiano}). In the static gauge
we identify the minkowskian coordinates of $ds^2_{Mink}$ in
eq.(\ref{metrik}) with the brane world-volume coordinates. The resulting induced
metric is
\be
ds_i^2=h^{-1/2}\left[-\left(1-h\vec{r}{\phantom{z\!}}'^2\right)d{\eta}^2+ d\vec x\cdot d\vec x\right]\ ,
\label{induced}
\ee
where a prime $(')$ denotes a derivative with respect to $\eta$ and $\vec r$ indicates the position of the brane  in the bulk.
In the case of slow (adiabatic) brane motion ($h {\vec{r}}{\phantom{z\!\!}}'^2\ll 1$) the brane action (\ref{actionn})
turns out to be
\bena
S&=&
\frac{T_3}{2}
\int d^4x \left(
r'^2+
r^2\Omega_5'^2
\right)\ . \label{dbikt}
\eena
In (\ref{dbikt}) we used
the symmetries of the background to write $\vec r'^2=r'^2
+r^2\Omega_5'^2$, where $\Omega_5$ parameterizes one of the ${\mathbb T}_{1,1}$ angles.

The equations of motion of the probe brane follow from varying the action (\ref{dbikt}) with respect to $r,\Omega_5$
\be
r''-r\Omega_5'^2=0\, , \ \ \ \ \ \ \ (r^2\Omega_5')'=0\ . \label{3}
\ee
First integrals of this system are provided by
\be
r'^2+\frac{J^2}{r^2}= 2 U \, , \ \ \ \ \ \
r^2\Omega_5'=J\ ,
\label{enn}
\ee
and eq.(\ref{3}) is then solved by the bouncing solution
\be
r=\sqrt{2U\,\eta^2+\frac{J^2}{2 U}}\, ,
~~~~~\Omega_5=\arctan\left(\frac{2U}{J}\,\eta\right)\, .\label{8}
\ee
A constant of integration has been fixed by requiring that
at $\eta=0$ the probe is at the turning point
$r_{min}={J}/{\sqrt{2U}}$. For discussions on
the exact solutions of (\ref{actionn}), one may see \cite{sling,AK,branonium,Kiritsis:2003mc,Brax:2002qw,gregory}.

One can easily find that the non-relativistic approximation is accurate whenever
${J^4}\gg{8 U^3}L^4$. In this approximation, the induced metric on the brane reads
\be
ds^2_i= h^{-1/2}ds^2_{\mbox{\tiny Mink}}\ .\label{metric}
\ee
An observer on the brane will therefore experience a
Friedman-Robertson-Walker metric with scale factor
\be\label{scale}
a(\eta)=h^{-1/4}={r}/({L \ln^{1/4}(r/r_s)})\, ,
\ee
 where $\eta$ is the
observer conformal time. Since $r(\eta)$ has a turning point, it is
easy to see that the same happens to $a(\eta)$, generating a
nonsingular bouncing cosmology.

The model is completed by  smoothly pasting this Mirage era to
a local gravity driven late evolution when the brane reaches the
top of the CY and gravity becomes localized {\em \'{a} la}
Randall-Sundrum \cite{rs}. There, the standard late time evolution of the observed Universe, is supposed to be well reproduced by the brane dynamics.
This assumption involves a transition from a mirage dominated era with a moving brane without any matter,
into a local gravity dominated era with a static brane and matter fields excited on it. This transition
has to be understood as an analogous of the reheating process in standard inflationary models. It entails
a dynamical mechanism under which the kinetic energy of the brane is passed to matter fields
and the brane motion is stabilized by acquiring a mass for the radion and and the angular scalar fields.
 The exact
description of this dynamics as well as the robustness of our predictions for physical observable is an
open point of the model which is left for future research.

\subsection{The Standard Cosmological Problems}
In \cite{sling} we showed how the standard cosmological problems are solved for the simpler version of the Slingshot
scenario, in which the probe brane moves in an $AdS_5$ throat. In what follows, we will complete the proof, extending it
to the KS case. An important ingredient in our argument is that the scale
factor for a brane moving in a KS throat (\ref{scale}) can be rewritten as a conformal re-scaling of the corresponding
scale factor for a brane moving in $AdS_5$, namely $a_{AdS}= r/L$. Indeed, we can write  (\ref{scale})  as
\be
a = \Omega_{ KS}(a_{AdS})\,a_{AdS} \ \ \ \ \ \ \ \ \mbox{with}\ \Omega_{ KS}(a_{AdS}) = \log^{-1/4}(a_{AdS}L/r_s)
\label{ads}\ .
\ee
%
Note that the above scaling should not be considered as a change of frame but just as a relation between two different
Friedman-Robertson-Walker scale factors (for example the Lagrangian will not be re-scaled as one would do for a conformal transformation).
It should be kept in mind that our approximations are valid whenever $a_{AdS}\gg r_s/L$.
Under such a re-scaling, the Hubble constant changes as
\be
H = \left(1+a_{AdS}\frac{d\ln\Omega_{ KS}}{da_{AdS}}\right)H_{AdS}
\label{hubblee}\ .
\ee
$H_{AdS}$ is the mirage Hubble constant of a brane moving in an $AdS_5$ throat \cite{sling}, namely
\be
H_{AdS}^2=-\frac1{L^2}\left[\frac{\kappa}{a^2_{AdS}}-\frac{2U}{a^4_{AdS}}+\frac{(J/L)^2}{a^6_{AdS}}+\frac{\kappa (J/L)^2-U^2}{a^8_{AdS}}\right]\ .
\label{hub}
\ee
We are now ready to study how standard cosmological problems are solved in the Slingshot scenario.

\vskip.3cm
\label{intro}
{\noindent\bf{A. Horizon.}} As explained in \cite{sling}, in the KS throat
the probe brane experiences a bounce in the String frame. This immediately ensures that
horizon problem is solved. To check this explicitly, we write the comoving horizon as
\be
\Delta\eta = \int_{\eta_i}^{\eta_0}d\eta
\ee
where $\eta_i$ is the smallest conformal time. To solve horizon problem it is required that $\Delta\eta > H_0^{-1}$.
Since we have $\eta_i\to-\infty$ due to the bounce, this condition is trivially satisfied.\vspace{.1cm}

\vskip.3cm
{\noindent \bf{B. Isotropy.}} In the $AdS$ case, mirage matter contributes to Hubble equation with a term
$\rho\sim a^{-8}_{AdS}$ \cite{sling}. This term dominates over the shear $\rho_{shear}\sim a^{-6}_{AdS}$ at early times,
avoiding the chaotic behavior \cite{Erickson:2003zm}.
To check whether this is true in the KS case, we should verify that the corresponding
mirage contribution dominates over the shear.
The form of this contribution can be read from (\ref{hubblee}). On the other hand,
the $\rho_{shear}$ term will still be given as $a^{-6}$ in the KS throat.
To compare them we can write the quotient as
\be
\sqrt{\frac{\rho_{shear}}{\rho}}\propto\left(1+a_{AdS}\frac{d\ln\Omega_{KS}}{da_{AdS}}\right)^{-1}\frac {a_{AdS}}{\Omega_{ KS}^3}\ .
\label{shear}\ee
The proportionality constant in (\ref{shear}) parameterize the anisotropic perturbations in the pre-bounce era.

\begin{figure}[t]
\centering
\includegraphics[angle=0,width=4.5in]{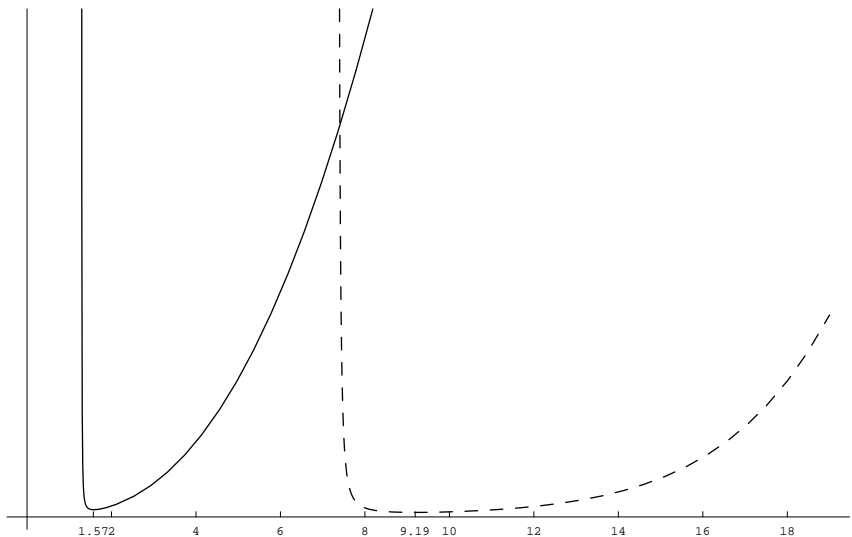}
\caption{
\label{isotropys}
The function $\frac{\rho_{shear}}{\rho}$ in the
String frame (solid line) and Einstein frame (dashed line) as a
function of $r/r_s$. The vertical axes has been plotted up to a proportionality constant that
can be adjusted in order to have small shear today. We note that, since our Universe brane moves in regions well to
the right of the horizontal axes $r\gg r_s$, shear never dominates.}
\end{figure}

It is simple to check that (\ref{shear}) is an increasing function of $a_{AdS}$ in the region $a_{AdS}\gtrsim1.57 \,r_s/L$ (figure \ref{isotropys}).
As we assumed that the Slingshot brane never approaches the tip of the KS throat, this condition is automatically satisfied.
Therefore, ${\rho_{shear}}/{\rho}$ decreases very rapidly close to the bouncing point in the pre-bounce era solving the isotropy problem.

\vskip.3cm
{\noindent \bf{C. Flatness.}} The curvature contribution to the Hubble
equation \footnote{The details of how to include a curvature term in the mirage Hubble equation can be found in \cite{sling}.}
can be disregarded if the quantity $|\Omega_{\mbox{\tiny Total}}-1|\,=\,{1}/{a^2H^2}$ passes through a minimum where it satisfies the
phenomenological constraint
\begin{equation}
|\Omega_{\mbox{\tiny Total}}-1|_{min}<10^{-8}
\, ,\label{cons}
\end{equation}
where the value $10^{-8}$ represents the measured curvature during the BBN.

In the $AdS$ case, this condition results in a restriction to a two dimensional region of parameter space.
In this sense flatness problem might be alleviated in the Slingshot scenario.

For the KS case we have, after conformal re-scaling
\begin{equation}
|\Omega_{\mbox{\tiny Total}}-1|=\frac {f^2}{a_{AdS}^2H_{AdS}^2}\ ,
 \ \ \ \ \ \ \ \
 \ \ \ \ \ \ \ \
 f=\frac{4\ln(a_{AdS} L/r_s)}{4\ln (a_{AdS} L/r_s)-1}\ .
\label{f1}
\end{equation}
The KT approximation is valid for $r_{min}\gg r_s$; to fix ideas we will use $r_{min}>10^2\ r_s$. In this region we have $f={\cal O}(1)$ and
decreasing in $a_{AdS}$. Consequently,
the flatness problem in the KS space might, in good approximation,
be alleviated by the same choice of parameters used in the $AdS$ case.

\section{Primordial Perturbations}
\label{sPert}
Before calculating the power spectrum of density perturbations, let us recall a lesson from inflation. In
inflation the primordial perturbations are produced by quantum
fluctuations of a scalar field, the inflaton. These fluctuations are
codified into the two point correlation function of the inflaton in
its vacuum state (the Bunch-Davis vacuum), which also sets the initial
conditions. However, these fluctuations are over-damped by the expansion of the Universe at super-horizon scales.
At these scales then, the quantum state becomes characterized by a large occupation number and
the system collapses into a classical state. This classical state
represent a random (gaussian) spectrum of perturbations
with variance given by the quantum correlations at the
quantum to classical transition point \cite{liddle-lyth} (see \cite{bozza} for other cases).

Let us now turn our attention to the mechanism proposed by \cite{wald}.
A perturbation of wavelength $\lambda$
smaller than the typical quantum scale (say $l_c$) of a given system,
is in its pure quantum state (vacuum). However, in an expanding background, the
wavelength of a perturbation grows in time ($\lambda\propto a$). In this
case whenever $\lambda\sim l_c$,
or in other words, as soon as the perturbation becomes macroscopic, wavelengths
bigger than the horizon scale collapse into a classical random state, with the same mechanism discussed before.
In the proposal of \cite{wald}, the relevant fluctuations are so continuously ``created''
at ``super-horizon'' scales.
Thus, a coherent (gaussian) spectrum of classical perturbations is produced
with variance given by matching the classical correlations with the quantum correlations
at the quantum to classical transition point \footnote{For a similar mechanism in the context of space-time non-commutativity see \cite{Mazumdar}.}.

In the Slingshot the mechanism of \cite{wald} is used, however the region in which the perturbations are frozen is not parameterized by the
horizon but by the centrifugal barrier, as we shall discuss later on.

Before working out the perturbation spectrum, a clarification must be added.
In the original proposal of \cite{wald} the perturbation was
produced by the same radiation which sets the Cosmic Microwave
Background. However, as pointed out by \cite{muk}, the perturbation coming out from the horizon today, should have been
born when the curvature sourced by the energy density of
radiation was much bigger than the Planck scale, which makes the
mechanism unreliable. In the Slingshot this problem is avoided.
There, perturbations are in fact created by brane fluctuations in a regime
in which the supergravity approximation is still valid. Moreover, the ``quantum gravity'' scale in this context, {\it i.e.} the string length,
is much smaller than the wavelengths of the relevant observable brane fluctuations.
For this reason, no extra quantum gravity effect
participate to the primordial perturbation spectrum.
 Finally, also note that the general result that bouncing cosmologies produce a blue spectrum in a Bunch-Davis vacuum
(see \cite{Brandenberger:2007by} for this result in mirage models) is avoided if the vacuum of \cite{wald} is used.

Having set the scenario, let us make more explicit the calculation of the primordial power spectrum.
In the mechanism introduced,
we can say that classical modes are created at the time $\eta_*$
when the proper wavelength of the corresponding quantum mode reaches the value
\be
a(\eta_*)/k\equiv a_*/k = l_c\, . \label{lc}
\ee
In the Slingshot scenario, a suitable value for the collapse length $l_c$ compatible
with all the observational constraints is $l_c=l_s g_s^{-\gamma}$ where $\gamma>1/3$ \cite{sling}.
Even if  this choice is not essential for the model, let us point out that is compatible
with a flat spectrum only in a background with a constant $g_s$, like KS or $AdS_5\times S_5$.

We start by perturbing the embedding of the probe brane by writing $r = r(\eta)+\delta r(\eta,\vec x)$ and
$\Omega_5 = \Omega_5(\eta)+\delta \Omega(\eta,\vec x)$.
The action (\ref{actionn}) can be expanded to quadratic order in $\delta$'s and their derivatives, getting in Fourier space
\bena
\!\!\!\!\!\!\!\!\!\!\!\!\!\!\!\!\!\!\!\!S=\frac{T_3}{2}\sum_k\int d\eta \left(
\phantom{\frac12}\!\!\! {\delta r}_k'^2 + r^2\delta {\Omega}'^2_k
-\left( k^2-\frac{J^2}{r^4} \right) \delta r_k^2 -
r^2k^2\delta\Omega_k^2 + \frac{4J}r\,\delta {\Omega}_k'\delta r_k
\right) .
\label{uno}
\eena
In what follows, we will find convenient to use the Bardeen potential \cite{durrer}
$\delta \Phi_k = \delta r_k/r$ instead of $\delta r_k$. With this change of variable we get
\be
\!\!\!\!\!\!\!\!\!\!\!\!\!\!\!\!\!\!\!\!\!\!\!\!\!\!\!\!\!\!S=T_3\sum_k\int
d\eta \left( \phantom{\frac12}\!\!\! \frac{r^2}2\left({\delta
\Phi'}_k ^2 + \delta {\Omega}'^2_k -k^2( \delta\Phi_k^2
+\delta\Omega_k^2) \right)+ J\,\delta {\Omega}_k'\delta \Phi_k
-J\,\delta {\Omega}_k\delta \Phi'_k \right) . \label{dos} \ee
The resulting generalized momenta are
\bena
&&\Pi_{\delta\Phi_k}=T_3\left(r^2 \delta \Phi'-J \delta \Omega\right)\, ,\\
&&\Pi_{\delta\Omega_k}=T_3\left(r^2 \delta \Omega'+J \delta \Phi\right)\, .
\eena
Then the Hamiltonian obtained form the above Lagrangian is
\be
\!\!\!\!\!\! H=\frac1{2T_3r^2}\left((\Pi_{\delta\Phi_k}+J\delta\Omega_k) ^2 +
(\Pi_{\delta\Omega_k}-J\delta\Phi_k) ^2\right) +\frac{1}2k^2r^2(
\delta\Phi_k^2 +\delta\Omega_k^2)\, ,  \label{H}
\ee
which is  positive definite  and has a well defined  ground state so that a
 quantum mechanical description of this system is possible.

\subsection{Normalized operators}

The  equations of motion for the fluctuations are derived from the action (\ref{dos}) and are written as
\bena \label{pertII}
\frac{d~}{d\eta}\left(r^2\delta\Omega'_k+2J\delta
\Phi_k\right)+r^2k^2\delta\Omega_k=0\ , \label{p1}
\\
\frac{d~}{d\eta}\left(r^2\delta\Phi'_k-2J\delta
\Omega_k\right)+r^2 k^2\delta\Phi_k=0\ ,\label{p2} \eena
The normalized solution for these equations are
\bena
 &&\delta\Phi_k=u_1 a_1+u_2 a_2+c.c.\\
 &&\delta\Omega_k=v_1 a_1+v_2a_2+c.c.\ ,
\label{quant}
\eena
where
\bena
 &&\!\!\!\!\!u_1=\sqrt{\frac{U}{k T_3}}\, \frac{\eta}{r^2}e^{-ik\eta}\, , ~~~~~~~~~~~~u_2=\sqrt{\frac{1}{U k T_3}}\, \frac{J}{2r^2}e^{-ik\eta}\\
 &&\!\!\!\!\!v_1=u_2=\sqrt{\frac{1}{U k T_3}}\, \frac{J}{2r^2}e^{-ik\eta}\, ~~~v_2=-u_1=-\sqrt{\frac{U}{k T_3}}\, \frac{\eta}{r^2}e^{-ik\eta}\ .
\label{quantII}
\eena
They satisfy the commutation rules
\bena
&&[\delta\Phi_k,\Pi_{\delta\Phi_{k'}}]=i\delta_{k,k'}\, ,~~~[\delta\Phi_k,\Pi_{\delta\Omega_{k'}}]=0\, ,
~~~\,[\delta\Omega_k,\Pi_{\delta \Phi_{k'}}]=0\, ,\\
&& [\delta\Omega_k,\Pi_{\delta\Omega_{k'}}]=i\delta_{k,k'}\, , \,
~~~~[\delta\Phi_k,\delta\Omega_{k'}]=0\, ,
~~~[\Pi_{\Phi_k},\Pi_{\delta \Omega_{k'}}]=0 \ ,\eena provided that
the operators $a_i,a_i^\dag$ are standard annihilation and creation
operators, {\it i.e.}, \be [a_i,a_j^\dag]=\delta_{ij}\, ,
~~~[a_i,a_j]=0\ . \ee
We are interested in the correlation of the Bardeen potential $\delta \Phi$ at the time of creation $\eta_*$. It
is straightforward to check that
\be
\langle\delta\hat\Phi_k\delta\hat\Phi_{k'}\rangle=\delta_{k,k'}\frac{1}{2kT_3r_*^2}\ ,
\ee
where $r_*=r(\eta_*)$.

\subsection{Classical Solution and Matching}

We will consider the transition  point of the quantum to the classical description in the region in which
$k\ll J/r^2$ (this was called the frozen region in \cite{sling}). Here in fact the oscillations of
the classical solutions of the equations of motion
are drastically damped and therefore the system can be considered classical, as it happens in inflation
\cite{liddle-lyth}. In this limit, we can discard the $k^2$ term in the equations (\ref{p1},\ref{p2})
and the resulting equations of motion read
\bena
\frac{d~}{d\eta}\left(r^2\delta\Omega'_k+2J\delta
\Phi_k\right)=0\ ,
\ \ \ \ \ \
\frac{d~}{d\eta}\left(r^2\delta\Phi'_k-2J\delta
\Omega_k\right)=0\ .\eena
The real solutions of these equations are
\bena\label{solutions}
&\delta\Phi_k&=\;\frac{C_k}{2J}+A_k\sin\left(2\theta+\phi_k\right)\
,\cr
&\delta\Omega&=-\frac{D_k}{2J}+A_k\cos\left(2\theta+\phi_k\right)\ ,
\eena
where $\phi_k,C_k,D_k,A_k$ are constants, $\theta=\Omega_5(\eta)-\Omega_5(\eta_*)$ and
\bena
C_k=r^2\delta\Omega_k'+2J\delta\Phi_k\ ,\cr
D_k=r^2\delta\Phi_k'-2J\delta\Omega_k\ .
\eena
With these definitions, it is easy to invert for $A_k$ to obtain
\be
A_k=\frac{r^2}{2J}\left[\delta\Phi_k'\cos\left(2\theta+\phi_k\right)-\delta\Omega_k'\sin\left(2\theta+\phi_k\right)\right]\ .
\ee
We now consider  initial conditions arising from the matching of the classical to the quantum system at the
time $\eta=\eta_*$. Therefore $C_k,D_k,A_k$ will be taken as Gaussian stochastic variables with
correlations $\langle...\rangle_c$ matching the quantum correlators $\langle...\rangle$ at $\eta=\eta_*$.
The constants $\phi_k$ will instead be related to the fact that the quantum
system picks up only positive frequencies in the vacuum state.

Using the quantum solutions described above at the matching point $\eta=\eta_*$ after a lengthly but
straightforward calculation we have
\bena \langle C_{k}
C_{k'}\rangle_{c}&=&\delta_{k,k'}\frac{k^2r_*^2+2U}{2kT}\ ,\cr
\langle A_{k}
A_{k'}\rangle_{c}&=&\delta_{k,k'}\frac{k^2r_*^2+2U}{8J^2kT}\ ,\cr
\langle A_{k}
C_{k'}\rangle_{c}&=&\delta_{k,k'}\frac{\sin\phi}{4kTJr_*^2}\{2J^2-2Ur_*^2-k^2r_*^4
-4JU\eta_*\cot\phi\}\ . \eena
The matching of
\be
\langle \delta\Phi_k\delta\Phi_{k'}\rangle_{c}=\langle \delta\hat\Phi_k\delta\hat\Phi_{k'}\rangle\ ,
\ee
requires $\phi_k=\pi/2$; this is the selection of positive frequencies.
So we are left for each mode with
\be
\delta\Phi_k=\frac{C_k}{2J}+A_k\cos2\theta\ .
\ee
In general correlators depend on time through $\theta$.
However in the frozen region, the oscillation rapidly stabilizes
in time. We will consider this asymptotic region ($2U\eta_{\mbox{\tiny asymp.}}/J>2\pi$) to be well before the nucleosynthesis. At this time
then
\be
\delta\Phi_k=\frac{C_k}{2J}-A_k\cos(2\Omega_*)=\frac{C_k}{2J}+A_k(1-2\frac{r_{min}^2}{r_*^2})\ .
\ee
Using the initial conditions found above we then get in the limit $k\ll J/r_*^2<J/r_{min}^2$
\be
\langle \delta\Phi_k\delta\Phi_{k'}\rangle_{c}\Big|_{\eta>\eta_{\mbox{ asymp.}}}\simeq \frac{\delta_{k,k'}}{2kTr_*^2}\left[1-
\left(\frac{r_{min}}{r_*}\right)^2\right]\ ,
\ee
so the power spectrum of temperature fluctuations is
\be
P(k)\simeq\frac{1}{2k\,T_3\,r_*^2}\left[1-
\left(\frac{r_{min}}{r_*}\right)^2\right]\ . \label{power}
\ee
A consistency condition for the production of the perturbation is that $r_{min}<r_*$. So we see that in the limit
$r_{min}\ll r_*$ we obtain the power spectrum introduced in \cite{sling}. Note that $r_*=r_*(k)$ as follows from eq.(\ref{lc}).

\subsection{Spectral index}

Since we assumed that perturbations are created
when the physical wavelength reaches a fixed value $l_c$, we have from eq.(\ref{lc}), $k l_c= a_*$. In the Klebanov-Tseytlin (KT) metric this means
\be
r_* = r_s \exp\!\left({-\frac14 \,W_{-1}(-\zeta)}\right)\, ,
\ee
where $\zeta=4r_s^4/L^4l_c^4k^4\leq e^{-1}$ and $W_{-1}(x)$ is the negative branch of Lambert's $W$-function
\cite{sling}. Then the power spectrum (\ref{power}) is explicitly written as
\be
P(k)=\frac{1}{2T_3\, k \, r_s^2} \,e^{{\frac12 \,W_{-1}(-\zeta)}}\!\left(1\!-\!\left(\frac{r_{min}}{r_s}\right)^2e^{\frac12 \,W_{-1}(-\zeta)}\right)\,,
\ee
whereas, the scalar spectral index $n_s=d\ln k^3P(k)/d\ln k$, turns out to be
\be
n_s = 1 + \frac{2}{1+W_{-1}(-\zeta)}\!\left(1-\frac{W_{-1}(-\zeta)}{1\!-\!\frac{r_s^2}{r_{min}^2}e^{-\frac12W_{-1}(-\zeta)}}\right)\,. \label{ns}
\ee
The first term in the parenthesis in (\ref{ns}) above is the redshift  due to the KT metric reported
in \cite{sling}. The second term in the parenthesis, gives a new correction coming form the time evolution of
the correlation function.
In the case in which $\zeta\ll 1$, we can use
the expansion of the Lambert W function for small argument $W(-\zeta)\simeq \ln(\zeta)+\cdots$ so to get
\be
P(k)\simeq\frac1{k^3 }\frac{1}{L^2l_c^2T_3}
\left(1-\left(\frac{r_{min}}{r_s}\right)^2 \!\!{\sqrt\zeta}\right)\,,
\ee
and
\be
n_s \simeq 1 + \frac{2}{\ln(\zeta)}-\frac{2\sqrt\zeta}{\sqrt\zeta-\frac{r_s^2}{r_{min}^2}}\,.
\label{hi}
\ee
Since the $\ln(\zeta)<0$ for small $\zeta$, the first correction on $n_s$ is negative.
On the other hand, the second correction comes from time evolution, and it is red or blue according to
the sign of its denominator. It will be negative whenever
\be
\sqrt\zeta>{r_s^2}/{r_{min}^2}\, ,
\label{hola}
\ee
from which we immediately see that long wavelengths are red-shifted.

On the other hand, if the last term is positive, then $\sqrt\zeta<{r_s^2}/{r_{min}^2}$ and the overall sign of the correction has to be evaluated
taking into account the joint contribution of both terms in (\ref{hi}). After some manipulations we find that the correction is red whenever
\be
{\sqrt\zeta}\left(1-2\,{\log\!\sqrt\zeta}\right)<\frac{r_s^2}{r_{min}^2}\, ,
\ee
from which we conclude that short wavelengths are also red-shifted, and there is an intermediate range of wavelengths that is blue-shifted.

The various parameters appearing in the formulas for the power spectrum and the spectral index may be partially fixed by
using the above expressions at a
pivot wavelength $\lambda_p=\zeta_p^{1/4}\lambda_0$ where $\lambda_0=a_0Ll_c/\sqrt{2}r_s$. Then we can write
\be
P(k_p)k^3_p\simeq\frac{1}{L^2l_c^2T_3}\left(1-\left(\frac{r_{min}}{r_s}\right)^2 \!\!{\sqrt\zeta_p}\right)= 10^{-10}\,,
\ee
and
\be
n_s \simeq 1 + \frac{2}{\log(\zeta_p)}-\frac{2\sqrt\zeta_p}{\sqrt\zeta_p-\frac{r_s^2}{r_{min}^2}}=.95\ ,
\ee
which gives two constraints for the three unknowns $r_{min}/r_s, L^2l_c^2T_3$ and $\lambda_0$.

Other constraints come from the requirement that the relevant primordial perturbations related to
the today's observational scales were born frozen.
These cross-constraints have been studied in \cite{sling} in the case in which $r_{min}\ll r_*$. The space of parameter is obviously larger
in the more general case in which $r_{min}< r_*$. However, for what concern the analysis in this paper we will be content in using the constraints
of \cite{sling}.

The last comment we need to add regards possible Trans-Planckian contributions to the primordial spectrum of perturbations.
As we have already stressed before, these contributions are not present in our model as the dynamics is always
controlled by the low energy regime of String Theory,
{\em i.e.} the supergravity approximation, and hence quantum stringy corrections are suppressed.
This is due to the fact that the brane is always slowly moving and that $r_{min}>l_s$, {\it i.e.} the brane is never too close to the stack.
The semiclassical treatment of the brane fluctuation is therefore the dominant effect.

\section{Four Dimensional Point of View}
\label{sEff}

The 4D effective theory for warped compactifications of IIB supergravity with (static) D-branes
has been derived by a perturbative approach in \cite{GM}, and
by a gradient expansion method in \cite{KK2}. In this last approach, the universal K\"{a}hler modulus $\rho(x)$
is obtained by writing the metric as \cite{KK2,KU}
\be
ds^2=h^{-\frac12}(x,y)\tilde g_{\mu\nu}(x,y)dx^\mu dx^\nu+h^{\frac12}(x,y)\gamma_{ab}(y)dy^a dy^b\ ,
\ee
with $h(x,y)=h(y)+h_0(x)$,  and identifying
\be
\rho(x)=iH(x)= i h_0(x) +\frac i{V_6} \int d^6 y \sqrt{\gamma} \,h(y) \ .
\ee
Here, $V_6$ is the volume of the (un-warped) compact manifold. The perturbative potential for $\rho$ has been
also obtained in \cite{KK2}
and it depends on the curvature of the transverse metric $\gamma_{ab}$, the localized negatively charged objects
like anti-branes and the $3$-form fluxes.
The effective four dimensional action is found to be \cite{KK2}
\be
S=\frac{1}{2\tilde k^2}\int d^4x \sqrt{-\tilde g}\left[HR[\tilde g]-2V(H)\right]\ ,\label{action0}
\ee
where $\tilde k^2=k^2_{10}/V_6$ and $k^2_{10}$ is the ten dimensional Newtonian constant.

We consider the background to be stabilized on the minimum of the potential of the universal K\"{a}hler modulus $V=0$
(by non-perturbative corrections or some other yet unknown mechanism) and we focalize on simplest case of an
$AdS_5\times S_5$ throat sourced by a stack of $N$ $D3$-branes, this can be easily generalized to the KS case.
In this case, the background close to the stack of $D3$-branes has the solution found by \cite{Gid} where $h_0(x)=0$,
$H\simeq 3{L^4}/{r_{max}^4}$ and  $r_{max}$ is a cut-off radius for which the
approximate solution of \cite{Gid} is no longer valid. Recalling that for this solution $L^4=4\pi l_s^4 N g_s$ and
plugging it into the action (\ref{action0}), we get an overall $N$ factor that can be reabsorbed by re-scaling the
metric $\tilde g_{\mu\nu}=N g_{\mu\nu}$ (and defining $k^2={r^4_{max}\tilde k^2}/{12\pi l_s^4 g_s}$).

We are now ready to add to the action (\ref{action0}) the contribution
of a moving brane. A single brane slowly moving on a background sourced by a stack of others $N$ D3-branes will
obviously produce gravity back-reactions sourced by the kinetic DBI energy of the brane.
In order to keep these back-reactions small we need to assume that we are in the adiabatic limit
$|h(y)g^{\mu\nu}\gamma_{ab}\partial_\mu y^a\partial_\nu y^b|\ll 1$.
The following effective action at order $1/N$ and in the slow velocity limit, {\it i.e.} at leading order
on the tension expansion in local curvature units, is thus obtained
\be
 S_{eff}
=\int d^4x \sqrt{-g} \left[\left(\frac{1}{2 k^2}-\frac{T_3L^2}{12N}\Phi^2\right)R-
\frac{T_3L^2}{2N}\left((\nabla \Phi)^2+\Phi^2(\nabla \Omega)^2\right)\right].
\label{eff1}
\ee
Here $\Omega$ and $\Phi=r/L$ parameterize respectively the angular and the linear motion of the moving D3 brane
on the warped CY. We have also included the conformal coupling to the four
dimensional Ricci scalar of \cite{grynberg}, which, consistently, is of order $1/N$.
The second term in (\ref{eff1}) is nothing else than the DBI action (\ref{actionn}) in the adiabatic limit.
In the expansion leading to (\ref{eff1}) the ratio $|h(y)g^{\mu\nu}\gamma_{ab}\partial_\mu y^a\partial_\nu y^b|/N$
has been considered to be at second order in the spirit of the Mirage approximation of \cite{KK}.

This action has been derived assuming a negligible backreaction on the CY geometry. This assumption should be valid
whenever the brane is deep inside the throat or,
in other words, whenever the value of the effective scalar field $\Phi$ (the radion) is small.
As soon as the brane exits the throat, {\it i.e.} for a large $\Phi$, the radion motion would acquire an effective (large) mass due to the
backreaction on the compact embedding. This in principle should stabilize $\Phi$ at late time. Together with that, another mechanism enters into
the stabilization of the effective String frame Newtonian constant. At late time $\Phi$ is supposed to couple to brane matter fields so to
re-heat the Universe.
From the ten dimensional point of view this re-heating can be understood as
the dissipation of the kinetic energy of the brane into world-volume excitations. As explained in the introduction, the details of this late
time dynamics are however out of the scope of the present paper and are left to future work.

\subsection{String frame}

The effective action (\ref{eff1}) does not however describe any physically relevant frame. In fact,
the metric $g_{\mu\nu}$ is not related to any physical choice of units. Units might be fixed either in terms of
particle masses (String frame) or of the four dimensional Planck length (Einstein frame).
In the String frame all particles masses are constant in time as
the Standard Model is supposed to live on the Slingshot brane. In the Einstein frame instead the gravitational
coupling is constant in time.
Although both the Einstein and String frames are of physical relevance, only the
String frame has a geometrical meaning from the ten dimensional
perspective. To make connection to what has been
described in the previous sections we will start discussing this case.

The induced metric in the slow velocity limit can be written as
\be
g_{\mu\nu}^i(x)=h^{-\frac12}(y(x))\left(g_{\mu\nu}(x)+h(y(x))
\gamma_{ab}(y(x))\,\partial_\mu y^a \partial_\nu y^b\right)\simeq
h^{-\frac12}(y(x))g_{\mu\nu}(x)
\, ,\
\ee
where $\partial_\mu y^a$ describe the brane embedding. In the simplest $AdS_5\times S_5$ case
$h^{-1/2}=r^2/L^2=\Phi^2$. The effective action on the String
frame is then
\be
\! \!\! \! \! \! \!\! \! \! \! \!\!\!\!\!\!\!\!\! S_{brane}=\int d^4x \sqrt{-g}\left[\left(\frac{1}{2k^2\Phi^2}-
\frac{T_3L^2}{12N}\right)R[g]+\frac{3}{k^2\Phi^4}(\nabla \Phi)^2-\frac{T_3L^2}{2N}(\nabla\Omega)^2\right]\ .
\label{eff2}
\ee
From the action (\ref{eff2}) we can finally find the effective Slingshot equations at zeroth order in the expansion.
Note that to get zeroth order back-reactions to the metric we need to consistently solve at first order the
scalar field equations.

At first order in $1/N$ expansion we have
\begin{eqnarray}\label{einstein}
G_{\alpha\beta}
&=&-2\frac{\nabla_\alpha\nabla_\beta \Phi}{\Phi}+
2\frac{\square \Phi}{\Phi}g_{\alpha\beta}-3\frac{(\nabla
\Phi)^2}{\Phi^2}g_{\alpha\beta}+
\\&&
+
\frac{T_3L^2k^2}{N}\left[\Phi^2\nabla_\alpha\Omega\nabla_\beta\Omega-\frac{\Phi^2}{2}(\nabla\Omega)^2g_{\alpha\beta}-
\frac{\Phi}{3}\nabla_\alpha\nabla_\beta \Phi+\frac{\Phi}{3}\square \Phi\ g_{\alpha\beta}-\frac{1}{2}
(\nabla \Phi)^2g_{\alpha\beta}\right]\ .\nonumber
\end{eqnarray}
By varying the action (\ref{eff2}) with respect to the
scalar fields $\Phi$ and $\Omega$ and by substituting the Ricci scalar obtained by the Einstein equations (\ref{einstein}),
the scalar field equations turn out to be, at leading order,
\begin{eqnarray}
&&\square\Omega=0\ ,\\
&&\square \Phi-2\frac{(\nabla \Phi)^2}{\Phi}=\Phi(\nabla\Omega)^2\ . \label{scalar}
\end{eqnarray}

\subsection{Cosmology}
\label{cosmology}

We now consider the theory (\ref{einstein}-\ref{scalar}) in a Friedman-Robertson-Walker (FRW) background. Here we have
\be
ds^2=-dt^2+a^2(t)d\vec{x}\cdot d\vec{x}\ ,
\ee
and all fields are  time dependent.

The equation for $\Omega$ reads
\begin{eqnarray}
\frac {d~}{dt}\left(a^3\dot \Omega\right)=0 \ \  \ \ \ \ \Rightarrow \ \ \ \ \ \ \ \dot\Omega=\frac J{a^3}\ ,
\end{eqnarray}
where $J$ is a constant of integration. Using this solution in the equation for $\Phi$ we find that
\begin{eqnarray}
\frac{\ddot \Phi}{\Phi}+3 \frac {\dot a}{a}\frac{\dot\Phi}{\Phi}-2\frac{\dot\Phi^2}{\Phi^2}=\frac{J^2}{a^6}\ .
\label{sscalar}
\end{eqnarray}
Moreover, from the $G_{tt}$ component of (\ref{einstein}), we get, at leading order in $1/N$, the Friedman equation
\begin{eqnarray}
\frac{\dot a^2}{a^2}&=&
\frac{\dot\Phi}{\Phi}
\left(2\frac {\dot a}{a}
-\frac{\dot\Phi}{\Phi}\right)\ ,\label{einsteinn}
\end{eqnarray}
which
is solved by the mirage solution $\dot a/a=\dot \Phi/\Phi$ or in other words by $a=C\Phi$. The constant $C$ can be fixed to
$C=1$ by appropriate
rescaling of the coordinates.
The scalar equation (\ref{sscalar}) ensures
instead the bouncing behavior
\begin{equation}
\frac{\ddot \Phi}{\Phi}=\frac{J^2}{\Phi^6}-\frac{\dot\Phi^2}{\Phi^2}\ ,
 \ \  \ \ \ \ \Rightarrow \ \ \ \ \ \ \
\dot\Phi^2= \frac{2U}{L^2\Phi^2}-\frac{J^2}{\Phi^4}\ ,
\end{equation}
where $U$ is another constant of integration and the bounce occurs at $\Phi^2=J^2/2UL^2$.
The effective theory therefore reproduces the bouncing solution of \cite{sling}
to leading order in the back-reaction parameters $|h(y)g^{\mu\nu}\gamma_{ab}\partial_\mu y^a\partial_\nu y^b|$
and $1/N$, which define the mirage approximation of \cite{KK}.

First order corrections are certainly interesting and even non-negligible when the brane leaves the CY
throat far from the bouncing point. However, the zeroth order solution  describes very well the motion of the Slingshot brane
deep into the throat of the CY. This is exactly the
framework in which the early time Slingshot is based.
We leave for future research the investigation of higher order corrections to the
mirage approximation.

\subsection{Energy conditions}

The Slingshot solution is ultimately related to the violation
of the energy conditions.  To verify this, let us define  the
effective energy condition $T_{\alpha\beta}$ as the right-hand side of
eq.(\ref{einstein}), which is, to leading order
\be
T_{\alpha\beta}=-2\frac{\nabla_\alpha\nabla_\beta \Phi}{\Phi}+
2\frac{\square \Phi}{\Phi}g_{\alpha\beta}-3\frac{(\nabla \Phi)^2}{\Phi^2}g_{\alpha\beta}\,.
 \ee
In a FRW background we may define the effective energy density $\rho=-T^t_t$ and the effective pressure $p\delta^i_j=T^i_j$.
By using the field equations (\ref{sscalar}-\ref{einsteinn}) we get
\be
\rho+3p=6H^2\left(1-\frac{\dot\Omega^2}{H^2}\right)\ ,
\ee
and
\be\label{p+r}
\rho+p=4H^2\left(1-\frac{\dot\Omega^2}{2H^2}\right)\ .
\ee
When the centrifugal barrier double the Hubble expansion, or in other words, when $\dot\Omega^2>2H^2$, both
strong ($\rho+3p\geq 0$) and weak ($\rho+p\geq 0$) energy conditions are violated and the bounce occurs. Strong energy conditions alone are however
violated as soon as the centrifugal barrier overtake the Hubble expansion, {\it i.e.} for $\dot\Omega>H$, this generates a short period of acceleration.
The number of e-foldings (${\cal N}$) of cosmological acceleration, can then be easily bounded by
\be
{\cal N}=\int H dt<\int \dot \Omega dt<2\pi\simeq 6\ .
\ee
A more precise numerical calculation shows that the number of e-foldings is actually bounded by $2$ \cite{sling,gregory}.

On the explicit solution it is easy to check that the energy conditions are both violated for
\be
a^2<\frac{3J^2L^2}{4U}\ .
\ee
From a more geometrical perspective,
brane curvatures violate energy conditions via the projection of bulk curvatures onto the brane, as discussed in \cite{sling}.

\subsection{Perturbations}

In section III we considered the primordial spectrum of perturbation by explicitly neglecting the local gravity contribution (mirage
approximation).
One might ask whether this is appropriate. Unfortunately, to fully answer this question a non-linear effective four
dimensional theory, in the expansion parameter $1/N$, must be developed.
Nevertheless, a convincing positive answer can be obtained by just looking at the
first order effective theory (\ref{einstein}). We will use the short notation $T_{\alpha\beta}$ for the right hand side of (\ref{einstein}).
Standard arguments \cite{dodelson}, shows that the local gravity contribution can be neglected if the Bardeen potential $\Psi$ satisfies
\be
\Psi\ll \frac{\delta T^t_t}{p+\rho}\ ,
\ee
whenever $\Psi$ is in the frozen region. In our model, all perturbations are frozen at their value $a_*\propto k$.
For small $k$s ({\it i.e.} for super-horizon scales) the Bardeen potential $\Psi$ is then well approximated with \cite{dodelson}
\be
3a_*^2H_*^2\Psi\simeq \frac{\delta T^t_t}{2}a_*^2\ ,
\ee
where $\delta T^t_t$ is the perturbation of the energy momentum tensor $T^t_t$.

The condition for which local gravity can be neglected in the calculation of the Bardeen potential is therefore
\be \label{ine}
\frac{1}{3}\ll \frac{1}{2}\frac{1}{1-\frac{a_b^2}{a_*^2}}\ ,
\ee
where we have used (\ref{p+r}). The inequality (\ref{ine}), is then always satisfied for large
scale modes.
This analysis therefore shows that local gravity can be safely neglected in the calculation of the scalar primordial perturbations related with the
CMB.

As we have shown in section \ref{cosmology}
when local gravity can be neglected, the effective action (\ref{eff2}) faithfully reproduce the mirage cosmology results.
We can define an effective matter content and pressure by making use of Hubble equation ${\dot a^2}/{a^2}=(8\pi G/3)\rho_{eff}$
and Raychaudhuri equation ${\ddot a}/{a}= -(4\pi G/3)(\rho_{eff}+3p_{eff})$. Since far form the bounce the angular momentum contribution
is subdominant, equation (\ref{sscalar}) imply
\begin{eqnarray}
\frac{\ddot a}{a}+\frac {\dot a^2}{a^2}= \frac{4\pi G}3\left(\rho_{eff}-3p_{eff}\right)\simeq0\ .
\label{raychard}
\end{eqnarray}
or in other words $p_{eff}=1/3 \rho_{eff}$. This is the well known result that mirage matter behaves like radiation \cite{AK}.
The Hollands-Wald result \cite{wald} of an almost flat power-spectrum can then be qualitatively
used from the effective theory point of view. Note however
that here perturbations are created whenever the angular momentum plays a determinant role. Nevertheless, the angular momentum signatures
on the later time evolution of the perturbations are expected to ``decay'' very quickly,
as shown before in the Mirage analysis.

\subsection{Einstein frame}

The Slingshot was originally developed in the so called String
frame \cite{sling}. In the String frame, particle masses are
defined by the oscillations of the fundamental string, and
therefore are constant in time. However, a  brane observer
experiences an induced time varying gravitational coupling until,
at late times, the brane leaves the throat and ends into the CY.
In \cite{Linde}, it
was claimed that a physical description of a gravitational system
may only be performed in a frame in which the induced
gravitational coupling is constant, {\it i.e.} in the so called
Einstein frame. However, as it is clearly explained in
\cite{veneziano}, the only requirement of the constancy of the
newtonian coupling ($G_4$) is meaningless. In local gravity for example, a physical meaningful
quantity is in fact not the Newton constant but rather the
gravitational force. In particular, the gravitational attraction
$F$  between two bodies of masses, let us say, $m_1$ and $m_2$, is
$F\propto G_4(t) m_1 m_2$. The change to the Einstein frame in
which $G_4=const.$, does not make the gravitational force
time-independent, as in this case masses are $m_i=m_i(t)$
\cite{veneziano}. The same conclusion can be obtained by considering
cosmological observations that involve the redshift.
The redshift is indeed defined as $z=\omega_1/\omega_2-1$ where $\omega_1$ and $\omega_2$ are respectively the
frequencies emitted and observed
of some atomic transition. In this physical system $\omega_1\propto m_e/a(t)$, where $m_e$ is the electron mass setting the units.
By a conformal transformation
$g_{\mu\nu}\rightarrow \Omega^2 g_{\mu\nu}$, we also have $m\rightarrow \Omega m$
and therefore the redshift is un-modified.
Finally, we also wish to quickly comment on the power spectrum of primordial perturbations.
In String frame the perturbed metric is $ds^2_S=a^2(1+2\Psi)dt^2+...$ where $\Psi$
is the Bardeen potential. By re-scaling to a conformal frame such that $ds^2_C=(a\Omega)^2(1+2\Psi)dt^2+...$,
we still obtain the same Bardeen potential $\Psi$. The power spectrum of primordial perturbations
(a physical quantity) is therefore unchanged
by the change of conformal frames.
To conclude, one can then in general show that any conformal frame
is physically equivalent \cite{faraoni}.

It seems much more economical and natural
to attribute the time dependence of the gravitational force to a
corresponding time variation of the gravitational coupling rather
than to the masses of the string excited states. Only during nucleosynthesis the question of whether the
gravitational coupling is constant or not is well posed and very
strict constraints do apply. Therefore, the time variation of the
four dimensional newtonian constant in the Slingshot era can never
be ruled out by observations.

During the Slingshot era, our Universe is a $D3$-brane
moving in a throat of a CY space. When the throat is embedded in a compactification scheme, it should be cut off at some $r_{UV}$
and glued back to the rest of
the compactification space \cite{Gid}. On the other hand, the conical singularity, at the tip of the cone,
is smoothed out by effectively
cutting off the throat at $r_{IR}$ and
appropriately deforming the space so that a smooth geometry is obtained. In the case in which the throat is taken to be $AdS_5\times S_5$ as
in \cite{sling}, and neglecting brane velocities contributions, this construction strongly resembles the RS1
scenario \cite{RS1} (see \cite{KK} and for a 10D tentative construction see \cite{KK2}), where
the vicinity of $r_{IR}$ and $r_{UV}$ correspond to the IR and UV branes, respectively (see fig.1).
\begin{figure}[t]
\centering
\includegraphics[angle=0,width=4in]{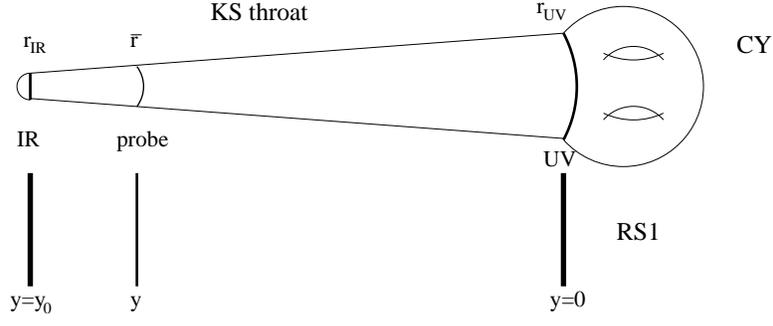}
\caption{The geometry of the KS throat glued to a CY. The tip of the throat is deformed at $r=r_{min}$ corresponding to an IR cutoff of the dual theory
whereas the "far" UV region of the throat is glued at $r=r_{max}$ to a CY space. The probe brane is wondering in the IR vicinity.
This is similar (in the zero angular momentum case) to the RS1 setup with a probe brane, where the tip corresponds to the IR brane
and the far region to the UV one.   \label{figu}}
\end{figure}
The correct analysis for
the induced Einstein theory on a third brane in the RS1 background has been performed in \cite{cotta}. The result for a probe brane is
\be\label{G}
8\pi G_4=\frac{2 \kappa_5^2}{L}\left[e^{2d_{UV}/L}-e^{-2 d_{IR}/L}\right]^{-1}\ ,
\ee
where $d_{IR}$ and $d_{UV}$ are the proper distances from the probe brane to, respectively, the $IR$ and $UV$
boundary and $\kappa_5$ is the higher dimensional gravitational coupling.
One can easily see that, once the probe brane reaches the $UV$ boundary, the newtonian constant stabilizes
to a constant value. In an exact $AdS_5$ geometry we have
\be
d_{IR}=\int^{r}_{r_{IR}} \frac{L}{r'}dr'=L \ln\left (r/r_{IR}
\right)\ ,
\ee
and
\be
d_{UV}=\int^{r_{UV}}_{r} \frac{L}{r'}dr'=L \ln \left( r_{UV}/r
\right)\ ,
\ee
where $r$ is the
probe brane position at fixed bulk time.
Thus we get that
\be
8\pi G_4=\frac{2 \kappa_5^2}{L}\left(\frac{r_{UV}^2}{r^2}-\frac{r_{IR}^2}{r^2}\right)^{-1}\ .
\ee
As the probe is moving in the vicinity of the IR region $r>r_{IR}$ and
far from the UV, $r\ll r_{UV}$, we may write (as already noted in \cite{KM})
\be
G_4\approx G_N\left(\frac{r}{r_{UV}}\right)^2
\ee
where $8\pi G_N\approx 2 \kappa_5^2/L$ is the Newton constant on the UV brane. Whenever instead the brane reaches the
UV boundary, the local gravitational coupling on the
moving brane stabilizes to the measured Newtonian constant.
The string theory system is slightly more involved as one needs to consider the conformal coupling between the local Ricci scalar
and the brane position introduced before. This is due to the fact that the bulk is not sourced by a cosmological constant, as in RS1 case,
but by a Ramond-Ramond five form. Nevertheless, we can use this toy model for a
qualitative description of what we would expect in the late time evolution of the Slingshot brane.

Let us consider two particles living on the brane. As the extra-dimensions are compact,
the Newtonian force between them behaves four-dimensionally,
{\it i.e.} $F\propto G_4/r^2$, at distances $r\gg \left(l_s^8/G_4\right)^{1/6}$.
When the probe brane reaches the UV region after the bounce, $G_4$
stabilizes to the measured newtonian constant and therefore
local 4D gravity dominates at low energies.
Viceversa, as the brane approaches the bouncing point, $G_4$
becomes smaller and smaller and therefore
the scale for which gravity looks four-dimensional, become
larger and larger. Eventually, gravity would look four-dimensional only above cosmological distances in the vicinity of the bouncing point.
In other words there, the $10D$ gravity dominates over the local gravitational attraction
up to very large scales and the probe
brane approximation of \cite{AK}
is justified. In this case, local $4D$ gravity cannot be used to describe gravitational attraction up to cosmological scales.

Although the effective theory discussed above is strictly valid only in the RS1 context, it actually
captures the essential fact that the gravitational coupling scales like $a^{-2}$ close to the IR boundary,
as pointed out by \cite{Linde}.
Let us indeed consider the String frame effective theory (\ref{eff2}), which is again valid
close to the bouncing point. The Einstein frame can be
found by re-scaling the metric by a conformal factor
\be
\chi^2=\frac{1}{\Phi^2}\left(1-\frac{T_3L^2 k^2\Phi^2}{6N}\right)\ .
\ee
The scale factor in Einstein frame is therefore
\be
a_E=\frac{a}{\Phi}\sqrt{1-\frac{T_3L^2 k^2\Phi^2}{6N}}=
1+{\cal O}\!\left(\frac{T_3L^2 k^2}{6N}\,\Phi^2\!\right)\ ,\label{ae}
\ee
where in the last equality we used the solutions of the equations (\ref{einstein}-\ref{scalar}), at leading order.
In Einstein frame, the spacetime is therefore not evolving at zeroth order \footnote{Nevertheless, a next to
leading order analysis of the background evolution
around the bounce show that $a_E$ evolves very slowly like radiation \cite{grandi}.}.
However, as already stressed before,
effective masses vary as $m_{eff}=\Phi^{-1} m$, where $m$ are the bare (String frame) particle masses. The physical
observable are therefore not only non-trivial in the Einstein frame, as conversely claimed by \cite{Linde},
but are physically equivalent to the physical observable in the String frame \cite{faraoni}.

To get an insight of what the next to leading order could be, we refer the reader to the exact solution of eqs.
(\ref{einstein}-\ref{scalar}) found in \cite{grandi}
\be
a=\frac1L\sqrt{\frac{J^2}{2U}+{2U}\eta^2}\ ,\ \ \ \ \ \ \Omega'=\frac{J}{L^2}\,\frac1{a^2}\,, \ \ \ \ \
\Phi=\frac{a}{1+k\sqrt{\frac{UT_3}{3N}}\,\eta}\ .\label{solu}
\ee
Note that in this solution there is a past singularity at negative $\eta$ in the field $\Phi$.
However the effective theory (\ref{eff2}) is valid for $\Phi^2\ll \frac{T_3L^2 k^2}{6N}$, {\it i.e.} the effective
theory cannot be trusted when the brane sits in the hat of the CY. Since from the ten dimensional point of view the
hat of the Calabi-Yau is a regular compact geometry, everything behaves regularly there and the value of $\Phi$ is
bounded from above by constant $\Phi_{max}$ determined by the details of the compactification.
In conclusion, the field singularity of solution (\ref{solu}) is only an apparent singularity and
if an exact solution could be found this should be regular in the past infinity and future. To properly show this
assertion however the full back-reacting Stringy solution must be performed, this is beyond the scope of this paper.

In terms of (\ref{solu}), the Einstein frame scale factor reads
\be
a_E=1+k\sqrt{\frac{UT_3}{3N}}\,\eta+{\cal O}\!\left(\frac{T_3L^2 k^2}{6N}\,\Phi^2\!\right)\ ,
\label{aaa}\ee
Note that the apparent past singularity on $\Phi$ gets mutated into a scale factor singularity,
however again the same discussions on its regularity apply.
The exact solution for the Einstein frame scale factor should then smoothly approach in the past infinity a constant
which is fixed by the scale of compactification. This resembles the emergent Universe idea \cite{EM}.
From the above considerations it is then clear that, even in the Einstein frame, no singularities should
appear whenever the Slingshot brane wanders in the throat of the CY, which is the case studied in this paper.

An interesting question is whether the spectrum of primordial perturbations analyzed in the probe brane approximation
can be reproduced from a four dimensional point of view. To achieve this goal,
a more detailed effective four dimensional theory than the one used here to describe the cosmological background, is needed. However, as
this study is far beyond
the scope of the present paper, we leave it for future work.

\section{Conclusions}
\label{sConcl}
The Cosmological Slingshot Scenario,
aspires to provide a new paradigm for the early time cosmology.
In there, the observable universe is a $D3$-brane moving in a
warped throat of a CY solution of the IIB supergravity. The inflationary era is
replaced by a period of Mirage Cosmology \cite{AK}, where local gravity effects are neglected.
At later times instead, when the probe brane is approaching
the base of the CY throat, local
gravity becomes important and completely takes over the dynamics.

In the Slingshot, the probe brane trajectory has a non-vanishing angular momentum in
the KS throat region of a CY space. The presence of the angular momentum provides a turning point on the
brane orbit at a finite distance to the tip of the throat.
From the point of view of an observer living in the brane, the
turning point prevents an initial singularity and gives rise to a
bouncing cosmology
 in String frame, without passing through a quantum gravity
regime. In Einstein frame on the other hand the scale factor starts as a finite constant in the past
corresponding to the Slingshot brane sitting at the hat of the Calabi-Yau.
In this context, by generalizing the results of \cite{sling} to a KS throat,
we showed that the problems of standard cosmology (horizon and isotropy) are naturally avoided and the flatness problem
is acceptably alleviated.

Concerning the calculation of the
quantum primordial perturbations, we have here identified their vacuum state
and elucidated the matching conditions from the quantum to the
classical regime of the perturbation. Using this detailed analysis, we have found the exact
power spectrum of primordial perturbations showing its compatibility with latest WMAP data \cite{wmap}
(see \cite{cri} for the exact match of the Slingshot primordial power spectrum with the full set of WMAP data).
A comment on a criticism raised by \cite{Linde} should be added here. As pointed out by \cite{muk} the scenario of \cite{wald} is un-reliable.
In \cite{wald} the primordial
perturbations are created by the same radiation that fills the Universe today.
However, if this is the case, the perturbation coming out from the horizon today would
have been born when the curvature scale was much bigger than the Planck scale. Fortunately, in the Slingshot this problem is avoided
as perturbations are created by brane fluctuations in a regime
in which the supergravity approximation is still valid.

In the second part of our paper we have studied the Slingshot cosmology from a four dimensional perspective.
In this context, we showed that the results found in \cite{sling}, obtained by using
the mirage approach of \cite{KK}, are reproduced.
Interestingly, we also clarified that strong and weak energy conditions are
violated in the effective theory approach during the bouncing.

In the Slingshot frame (or String frame), particle masses are fixed and Newton constant $G_4$ is time-dependent.
This time variation might
seem odd. However, no observational
constraints on the variation of $G_4$ are applicable to the pre-nucleosynthesis era.
The Slingshot, that aims for an early time description of the Universe,
is therefore observationally safe.
Nevertheless, a frame where $G_4$ is constant might be more familiar for some \cite{Linde}.
This frame, at leading order in the tension expansion with respect
local curvatures, is easily found from the action (\ref{eff2}) by a conformal transformation.
The same conformal transformation that fix in time the Newtonian constant, makes
however particle masses time varying. It is then easy to note that
the physical observable are frame-independent, as already elucidated by Dicke's
original paper \cite{Dicke:1961gz} (see \cite{faraoni} for a modern view).

Finally, one may ask of whether cosmological singularities might
occur in the Einstein frame.
Here we have showed that  during the early time cosmology, there
are no singularities in any frame. Although the effective theory
for the whole brane trajectory is not known, and thus, definite
statements cannot be made, one might speculate that  if no
singularities are formed during the brane motion deep into the
throat, future singularities, in the hat of the CY, are unlikely
to appear. The physical system reaches high energies only when the
brane is close to the stack. Therefore, from the ``mirage''
perspective, there cannot possibly be any future cosmological
singularities \footnote{Here we assume that there are no brane
collisions, embedding singularities in the hat of the CY and/or
singular dynamics of the CY volume.}. As Physics is frame
independent, the same conclusion should be valid in Einstein frame
as well. Nevertheless, a quantitative exploration of this point,
involving the transition from the mirage to the local era, should
be done in future research.

\acknowledgements

CG wishes to thank Valerio Bozza and Filippo Vernizzi for
enlightening discussions on cosmological perturbations. CG also wishes to thank Cliff Burgess for the discussion on physical observable
on different conformal frames. Finally, CG
wishes to thank the Physics Department of the King's College London
for the hospitality during part of this work. CG and AK wish too thank Elias Kiritsis for useful comments.
NEG thanks H. Vucetich
and G. A. Silva for helpful discussions on the definition of newtonian constant. CG and NEG wish to thank
Susha Parameswaran for discussions on possible realizations of the Slingshot scenario in the RS1 context.

\end{document}